\documentstyle[prl,preprint,aps]{revtex}
\begin{document}
\draft
\preprint{to appear in Phys. Rev. Lett.}
\title
{Analytic results for $N$ particles with $1/r^2$ interaction in\\
two dimensions and an external magnetic field}
\author{Neil F. Johnson$^1$ and Luis Quiroga$^2$}
\vspace{.05in}
%
\address
{$^1$Department of Physics, Oxford University, Oxford, OX1 3PU,
England}
\address
{$^2$Departamento de Fisica, Universidad de Los Andes, Bogota,
A.A. 4976,
Colombia}
%
\maketitle

\begin{abstract}
The $2N$-dimensional quantum problem of $N$ particles
(e.g. electrons) with
interaction
$\beta/r^2$ in a two-dimensional parabolic potential $\omega_0$
(e.g. quantum dot) and
magnetic
field $B$, reduces exactly to solving a
$(2N-4)$-dimensional problem
which is {\em independent} of $B$ and
$\omega_0$.
An exact, infinite set of relative mode excitations are obtained
for
any $N$. The $N=3$ problem reduces to that of a
ficticious particle in
a two-dimensional, non-linear potential of strength $\beta$,
subject to a ficticious
magnetic field $B_{\rm fic}\propto J$, the relative angular
momentum.

\end{abstract}
\vspace{1.0in}


\narrowtext

Few-body problems have always attracted interest in the fields of
atomic and
nuclear physics. Recent work on laser-cooled ions in Paul
traps \cite{diedrich} has heightened their
importance in atomic physics.
In condensed matter physics, such problems have been used
indirectly as cluster calculations for understanding
many-electron
systems such as the two-dimensional (2D) electron gas in a
magnetic
field.
A famous example is Laughlin's numerical calculation for $N=3$
electrons
in a 2D parabolic potential used for investigating the
Fractional
Quantum Hall Effect  \cite{laughlin}.
Few-body problems have recently taken on more direct relevance in
semiconductor
physics
due to rapid advances in fabrication of quantum dots containing
few
electrons  \cite{ashoori}  \cite{mceuen} \cite{Maksym}
\cite{Hawrylak}.
In lateral quantum dot structures, the electrons are typically
free to
move in only
two spatial dimensions and the confining potential is
approximately
parabolic  \cite{ashoori}  \cite{mceuen}.
A complete description of this few-electron system is
complicated
since the confinement energy, the
electron-electron repulsion and the cyclotron energy due to
applied
magnetic
fields are typically comparable in magnitude. Numerical
perturbative
approaches employing a basis of non-interacting single-particle
states
become
computationally intensive in the strongly-interacting
(Wigner solid)
regime.
Analytic simplifications of the exact $N$-particle Hamiltonian or
exact
solutions
of model $N$-particle Hamiltonians can therefore be useful.

Few-body Hamiltonians are rarely solvable analytically.
Exceptions include
$N$ particles in 1D with $\beta/r^2$
interaction \cite{calogero}
 and $N=2$ electrons
in 2D with
$\beta/r^2$ interaction \cite{quiroga} and magnetic field.
Here we show that the $2N$-dimensional problem of $N$ particles
(e.g. electrons) with $\beta/r^2$
interaction
in a 2D parabolic potential $\omega_0$ (e.g. quantum dot) and
magnetic
field $B$ reduces exactly to solving
a $(2N-4)$-dimensional problem which
is {\em independent} of $B$ and $\omega_0$. An exact set of
relative
mode
excitations
are obtained.  The $N=3$ particle problem reduces to that of a
particle moving in
a 2D non-linear potential of strength $\beta$, subject
to a {\em ficticious}
magnetic field $B_{\rm fic}\propto J$, the total relative
angular
momentum. The ground state $J$ (i.e. magic number) transitions
for $N=3$
are quantitatively consistent with
numerical
calculations for the Coulomb interaction\cite{Hawrylak}.
Analytic results are given in the Wigner solid regime.
The present work implicitly includes mixing with
all Landau levels.

The exact
Schrodinger equation for $N$ particles with repulsive
interaction $\beta/r^2$
moving in a 2D parabolic potential
subject to a magnetic field $B$ (symmetric gauge)
along the z-axis,
is given by $(H_{\rm space}+H_{\rm spin})\Psi=E\Psi$;
\begin{equation}
H_{\rm space}=\sum_{i=1}^{N} (\frac{{\bf p}_i^2}{2m^*}
+
{\frac{1}{2}}m^*\omega_0^2(B) |{\bf r}_i|^2
+ \frac{\omega_c}{2} l_i)+
\sum_{i<j} {\frac{\beta}{|{\bf r}_i-{\bf r}_j|^2}}
\end{equation}
where $\omega_0^2(B)=\omega_0^2+\frac{\omega_c^2}{4}$,
$\omega_c$ is
the cyclotron
frequency, and
$H_{\rm spin}=-{g^*\mu_B B}\sum_i s_{i,z}$.
The momentum and position of the $i$'th particle are
given by 2D
vectors ${\bf p}_i$ and ${\bf r}_i$ respectively; $l_i$
is the
z-component
of the
angular momentum.
The exact eigenstates
are written in terms of products of spatial and spin
eigenstates
obtained
from
$H_{space}$ and $H_{spin}$ respectively; eigenstates of
$H_{\rm spin}$ are just products of the spinors of the
individual
particles.
We employ standard Jacobi coordinates ${\bf X}_i$
($i=0,1,\dots ,N-1$)
where
${\bf X}_0=\frac{1}{N}\sum_j {\bf r}_j$ (center-of-mass),
${\bf X}_1=\sqrt{\frac{1}{2}}({\bf r}_2-{\bf r}_1)$,
${\bf X}_2=\sqrt{\frac{2}{3}}(\frac{({\bf r}_1+
{\bf r}_2)}{2}-{\bf r}_3)$
etc.
(see Fig. 1 for $N=3$)
together
with their conjugate momenta ${\bf P}_i$.
The center-of-mass motion decouples,
$H_{\rm space}=H_{\rm CM}({\bf X}_0)+
H_{\rm rel}(\{{\bf X}_{i>0}\})$,
hence $E_{\rm space}=E_{\rm CM}+E_{\rm rel}$.
The exact
eigenstates of $H_{\rm CM}$ and energies $E_{\rm CM}$
are well-known
\cite{fock}.
The non-trivial problem
is to solve
the relative motion equation
$H_{\rm rel}\psi=E_{\rm rel}\psi$.
We transform the relative coordinates
$\{{\bf X}_{i>0}\}$ to standard hyperspherical
coordinates:
${\bf X}_i=r(\prod_{j=i}^{N-2}{\rm sin}\alpha_{j+1}){\rm cos}
\alpha_{i}
e^{i\theta_i}$ with
$r\geq 0$ and $0\leq\alpha_i\leq\frac{\pi}{2}$
($\alpha_{1}=0$).
Because $J$ remains a good quantum number, we
introduce a Jacobi transformation of the relative motion
angles $\{\theta_i\}$:
in particular
$\theta'=\frac{1}{N-1}\sum_{i=1}^{N-1}\theta_i$,
$\theta=\theta_1-\theta_2$ etc.
(see Fig. 1 for $N=3$). We hence have $(N-1)$
$\theta$-variables,
$(N-2)$ $\alpha$-variables and a hyperradius $r$ giving a
total of
$2(N-1)$ variables as required for the relative motion.
The exact eigenstates of $H_{\rm rel}$ have the form
$\psi=e^{iJ\theta'}R(r)G(\Omega)$ where $\Omega$ denotes
the $(2N-4)$
remaining
$\{\theta,\alpha\}$ variables excluding $\theta'$;
$R(r)$ and $G(\Omega)$ are
solutions of the
hyperradial and $(2N-4)$-dimensional hyperangular
equations respectively.
The hyperradial equation is
\begin{equation}
\big(\frac{d^2}{d r^2}+
\frac{2N-3}{r}\frac{d}{d r}-\frac{\gamma(\gamma+2N-4)}{r^2}-
\frac{r^2}{l_0^4}+\frac{2m^*(E_{\rm rel}-
\hbar J\frac{\omega_c}{2})}{\hbar^2}
\big)R(r)=0
\end{equation}
where $l_0^2=\hbar(m^*\omega_0(B))^{-1}$; the parameter
$\gamma>0$ and is related to the
eigenvalue of the $B$ and $\omega_0$-independent
hyperangular equation (see below).
Equation (2) can be solved exactly yielding
\begin{equation}
E_{\rm rel}=\hbar\omega_0(B)(2n+\gamma+N-1)+
J\frac{\hbar\omega_c}{2}
\end{equation}
where $n$ is any positive integer or zero and
\begin{equation}
R(r)=\big(\frac{r}{l_0}\big)^\gamma L_n^{\gamma+N-2}
\big(\frac{r^2}{l_0^2}\big)
e^{-\frac{r^2}{2l_0^2}}\ \ .
\end{equation}
Equation (3) provides an exact (and infinite) set
of relative mode excitations $2\hbar\omega_0(B)\Delta n$
for any
$N$ {\em regardless} of particle statistics and/or
spin states.
These are ``breathing"
modes, as shown below for $N=3$;
numerical Coulomb results have shown similar modes
to this set of $\beta$-independent excitations \cite{review}.
All that remains is to solve
the
$B$ and $\omega_0$-independent hyperangular equation
which
resembles a
(single-particle)
Schrodinger-like equation
in $(2N-4)$-dimensional
$\Omega$-space.
The eigenvalue of the hyperangular equation
\begin{equation}
\epsilon=\frac{\hbar^2}{8}\big[\gamma(\gamma+2N-4)-J^2-
\big(\frac{V_{\rm class}}
{\hbar\omega_0(B)}\big)^2\big]
\end{equation}
where $V_{\rm class}$ is the potential energy of
the classical, minimum-energy
configuration (Wigner solid);
$V_{\rm class}\propto\beta^{\frac{1}{2}}\omega_0(B)$ and
hence $\epsilon$ (like $\gamma$) is independent of $B$ and
$\omega_0$.
The exact relative energy for any $N$
\begin{equation}
E_{\rm rel}=\hbar\omega_0(B)\big[2n+
\big([N-2]^2+J^2+\big(\frac{V_{\rm class}}
{\hbar\omega_0(B)}\big)^2+\frac{8\epsilon}
{\hbar^2}\big)^{\frac{1}{2}}
+1\big]\\
+J\frac{\hbar\omega_c}{2}\ \ .
\end{equation}
$E_{\rm rel}$ only depends on particle statistics
through
$\epsilon$.  As $\hbar\rightarrow 0$,
$E_{\rm rel}\rightarrow V_{\rm class}$
and $\epsilon\rightarrow 0$.
Physically, $\epsilon$ accounts for the
``zero point energy" in $\Omega$-space
associated with the quantum-mechanical spread of
$G(\Omega)$
about the hyperangles $\Omega$ corresponding to the
classical, minimum energy configuration
(Wigner solid); the actual spread in $G(\Omega)$
and hence $\epsilon$
will depend on
total wavefunction symmetry
requirements (see below for $N=3$).
In general $\epsilon\geq 0$,
$\epsilon\sim \beta^\mu$ where $\mu<1$ (the dominant
$\beta$-dependence of $E_{\rm rel}$
lies in
$(V_{\rm class})^2$) and $\epsilon\sim J^\delta$
where $\delta<2$.
Equation (6) implies that for any $N$,
the ground state $J$ value will tend to become
increasingly large and negative
with increasing
$B$-field ($\omega_c>0$, e.g. electrons).
We now demonstrate these statements
explicitly for $N=3$.

For $N=3$ we change variables from $\alpha,\theta$ to
$x,y$ where
$x={\rm ln}({\rm tan}\alpha)$ and
$y=\theta-\frac{\pi}{2}$. Since
$0\leq\alpha\leq\frac{\pi}{2}$, hence
$-\infty\leq x\leq\infty$ (N.B. $-\pi\leq y\leq\pi$).
We define
$p_x=\frac{\hbar}{i}\frac{\partial}{\partial x}$ and
$p_y=\frac{\hbar}{i}\frac{\partial}{\partial y}$.
The exact hyperrangular equation can be written in the
form
\begin{equation}
\big[\frac{p_x^2}{2}+\frac{(p_y+\frac{\hbar
J {\rm cos}(2{\rm tan}^{-1}e^x)}{2})^2}
{2}\\
+ V(x,y;\epsilon)\big]G(x,y)=\epsilon G(x,y)
\end{equation}
where
\begin{eqnarray}
 V(x,y;\epsilon)=m^*\beta\big[
\frac{(2+{\rm cos}(2{\rm tan}^{-1}e^x)}{({\rm cosec}
(2{\rm tan}^{-1}e^x)+
{\rm cot}({\rm tan}^{-1}e^x))^2- 3{\rm sin}^2 y}
- \frac{3}{4}{\rm sin}^2(2{\rm tan}^{-1}e^x)\nonumber\\
+\frac{1}{2}{\rm cos}^2({\rm tan}^{-1}e^x)+
\frac{\epsilon}{m^*\beta}{\rm
cos}^2(2{\rm tan}^{-1}e^x)\big]   \ \ .
\end{eqnarray}
Equation (7) represents the single-body
Hamiltonian for a ficticious
particle of energy
$\epsilon$ and unit mass, moving in the
xy-plane in a non-linear (i.e. $\epsilon$-dependent)
potential
$ V(x,y;\epsilon)$,
subject to a ficticious, non-uniform magnetic field
in the
$z$-direction
\begin{equation}
B_{\rm fic}= \frac{\hbar J c}{4e}\big[1-{\rm cos}
(4({\rm tan}^{-1}e^x))\big]\ \ .
\end{equation}
$B_{\rm fic}$ is independent of $B$ and
has a maximum of
$\frac{\hbar |J| c}{2e}$   at $x=0$ for all $y$.
For small $x$, $B_{\rm fic}\approx \frac{\hbar J
c}{2e} (1-x^2)$. As
$x\rightarrow\pm\infty$, $B_{\rm fic}\rightarrow 0$.
Note we have here chosen to highlight the
Schrodinger-like form of Eq. (7);
a simple rearrangement of Eq. (7)
shows it to be hermitian with a weighting function
${\rm sin}^2(2{
\rm tan}^{-1}e^x)$ \cite{referee}.
Our results are exact so far.

Figure 2 shows the potential $ V(x,y;\epsilon)$ in
the $(x,y)$ plane.
$V(x,y;\epsilon)\geq 0$ everywhere.
Minima occur at $(0,0)$ and $(0,\pm\pi)$ where
$V(x,y;\epsilon)=0$ (N.B. $(0,\pi)$ is equivalent to
$(0,-\pi)$).
Maxima occur at $({\rm ln} {\sqrt 3}, \pm\frac{\pi}{2})$
in Fig. 2,
where
$V(x,y;\epsilon)\rightarrow\infty$.
Since $\epsilon\geq 0$, these statements hold for any
$\epsilon$.
We now discuss the physical significance of these
features.
The classical configurations of minimum energy
(Wigner solid)
correspond to the
particles
lying on a ring in the form of an equilateral
triangle with
$V_{\rm class}=\omega_0(B)[6m^*\beta]^{\frac{1}{2}}$.
There are two distinct configurations with clockwise
orderings
$(132)$ and $(123)$
corresponding to
$(\alpha,\theta)=(\frac{\pi}{4},\pm \frac{\pi}{2})$.
In $(x,y)$
coordinates, these
correspond to $(0,0)$ and $(0,\pi)$ (equivalently,
$(0,-\pi)$).
Hence the classical configurations coincide with the
minima in
$ V(x,y;\epsilon)$ in Fig. 2 and the maximum in
$B_{\rm fic}$.
The formation of the Wigner solid should therefore be
favored by
both
large $B_{\rm fic}$ (i.e. large $|J|$) and deep
$ V(x,y;\epsilon)$
minima
(i.e. large $\beta$, strong electron-electron
interactions).

Consider the limit of three electrons with very
strong
electron-electron
interactions
(i.e. $\beta\rightarrow
\infty$).
Since the tunnel barrier height between the two
$ V(x,y;\epsilon)$ minima $\sim \beta$, the ficticious
particle
sits at one of these minima and the system is locked
in one of the two classical configurations, e.g. $(132)$
at
$(0,0)$.
The tunnelling probability between the minima is zero.
Tunnelling
between the
two minima implies a mixture of configuration $(123)$
into $(132)$ and hence interchange of the original
electrons; in
many-body
language exchange effects arising from wavefunction
antisymmetry
are therefore negligible.
$\epsilon$ is small compared to $m^*\beta$ and Eq. (6)
reduces to
\begin{equation}
E_{\rm rel}=\hbar\omega_0(B)\big[2n+
(1+J^2+\frac{6m^*\beta}{\hbar^2})^{\frac{1}{2}}
+1\big]+J\frac{\hbar\omega_c}{2}\ \ .
\end{equation}
The energy $E_{\rm rel}\geq
V_{\rm class}$ since it includes the hyperradial
zero-point energy
(N.B. $\hbar\rightarrow 0$
yields $E_{\rm rel}\rightarrow V_{\rm class}$ and
$B_{\rm fic}\rightarrow 0$).

Next consider large but finite $\beta$.
The ficticious particle now
moves in the vicinity of the minimum (i.e.
$(x,y)\approx (0,0)$).
The electrons in the Wigner solid are effectively
vibrating around
their
classical positions.
Expanding the potential $ V(x,y;\epsilon)$ about $(0,0)$
to third
order,
the exact Eq. (7) becomes
\begin{equation}
\big[\frac{p_x^2}{2}+\frac{(p_y-\frac{\hbar J x}{2})^2}{2}
\\
+\frac{1}{2}\omega_x^2x^2+\frac{1}{2}\omega_y^2y^2\big]
G(x,y)=\epsilon G(x,y)
\end{equation}
where $\omega_x^2=(\frac{3m^*\beta}{4}+2\epsilon)$
and
$\omega_y^2=\frac{3m^*\beta}{4}$. This has the form
of
a single electron moving in an anisotropic parabolic
potential,
subject to
a uniform
magnetic field $B_{\rm fic}=\frac{\hbar J c}{2e}$.
Equation (11) is exactly solvable for $\epsilon$
using a symmetric
gauge\cite{madhav} (the energies are independent
of the choice of
gauge for
$B_{\rm fic}$). A full discussion of the results
for any $\epsilon$
will be presented elsewhere. As an illustration,
we consider
small $\epsilon$ hence $\omega_x\approx\omega_y$.
Equation (6)
becomes
\begin{eqnarray}
E_{\rm rel}=\hbar\omega_0(B)(2n+
\big[1+J^2+\frac{6m^*\beta}{\hbar^2}+
2(2n'+|l|'+1)
(J^2+\frac{12m^*\beta}{\hbar^2})^{\frac{1}{2}}+2l'J
\big]^{\frac{1}{2}}
+1)\nonumber\\
+J\frac{\hbar\omega_c}{2}\ \ .
\end{eqnarray}
The ficticious particle has its own set
of Fock-Darwin (and hence Landau) levels \cite{fock}
labelled by
$n'$ and a
ficticious
angular momentum $l'$.
For large $\beta$ and small $n',l'$ and $J$, Eq. (12)
yields an
oscillator
excitation spectrum with two characteristic frequencies
${\sqrt 2}\hbar\omega_0(B)$ and $2\hbar\omega_0(B)$
representing
``shear" and
``breathing" modes of the Wigner solid.

For smaller $\beta$ (i.e. weaker interactions) and/or larger
$\epsilon$ (i.e.
excited states),
the
tunneling probability between the $V(x,y;\epsilon)$ minima
in Fig. 2 becomes significant; the Wigner solid
begins to melt and
wavefunction antisymmetry (exchange) must be included. For
three
spin-polarized electrons,
$\psi$ must be antisymmetric under particle interchange
$i\leftrightarrow j$.
The hyperradial part $R(r)$ is
invariant; particle permutation operations in
$({\bf r}_1,{\bf r}_2,{\bf r}_3)$
become
straightforward {\em space-group} operations in the
$(x,y)$ plane.
For small $(x,y)$, $1\leftrightarrow 2$
is equivalent to $(x,y)\rightarrow (x,y+\pi)$ with
$\theta'\rightarrow \theta'+
\frac{\pi}{2}$; $1\leftrightarrow 3$
is equivalent to $(x,y)\rightarrow (\bar x,\bar y-\pi)$
with $\theta'\rightarrow \theta'+
\frac{\pi}{6}$ ($(\bar x, \bar y)$ represents $(x,y)$
rotated by
$\frac{4\pi}{3}$); $2\leftrightarrow 3$
is equivalent to $(x,y)\rightarrow (\tilde x,\tilde y+\pi)$
with $\theta'\rightarrow \theta'-
\frac{\pi}{6}$ ($(\tilde x, \tilde y)$ represents $(x,y)$
rotated
by
$-\frac{4\pi}{3}$).
Single-valuedness of $\psi$ implies
$e^{\pm iJ\pi}G(x,y\pm 2\pi)=G(x,y)$.
Note we have implicitly satisfied Bloch's theorem in this
analysis, i.e.
$G(x,y\pm 2\pi)=e^{\pm i 2\pi k}G(x,y)$.
The solutions $G(x,y)$ of Eq. (7) with the lowest possible
$\epsilon$
(and hence lowest $E_{\rm rel}$ at a given $\omega_c$) should be
nodeless
in the vicinity of $(0,0)$
(c.f. ground state in the parabolic potential with $n'=0=l'$ in
Eq. (12)).
However the above symmetry requirements forbid such a nodeless
solution {\em unless}
$e^{i\pi\frac{2J}{3}}=1$. Therefore the only symmetry-allowed
solutions $G(x,y)$
which are nodeless (i.e. smallest $\epsilon$ and hence lowest
$E_{\rm rel}$ at a
given $\omega_c$)
are
those where $J$ is a multiple of three.
Evaluating the simplified expression for $E_{\rm rel}$ in Eq.
(12) ($n'=0$$=l'$),
the following ground
state $J$ transitions are obtained with increasing $\omega_c$
for
three spin-polarized electrons
in a GaAs
dot  ($\hbar\omega_0=3.37$meV as in
Ref. \cite{Hawrylak})\cite{beta};
$-3\rightarrow -6$ at $B=5.0$T,
$-6\rightarrow -9$ at
$B=8.7$T, and  $-9\rightarrow -12$ at $B=12.2$T (N.B. $J=0$
is not allowed by
symmetry).
The numerically obtained values from Ref. \cite{Hawrylak} are
$B\sim 5.5$T,
$8.4$T
and
$12.4$T using a Coulomb interaction. Our analytic
results therefore agree
well with the
numerical calculations despite the different interaction form
(see below).
A feature of these analytic results is that they become more
accurate in the
Wigner solid regime (e.g. large $\beta$ or $|J|$) while the
numerical calculations become more
computationally demanding.

For general $N$, the hyperrangular equation (c.f. Eq. (7))
becomes
$(2N-4)$-dimensional.
However in the Wigner solid regime (large $\beta$ or $|J|$)
the classical minimum energy configurations will still be
important in determining $\epsilon$
and hence $E_{\rm rel}$, just as for $N=3$.
The classical minimum energy configurations
(with $1/r$ interaction) for $N<6$ all
consist
of
$N$ particles on a
ring; for $N=6$ additional minima occur \cite{bolton}.
Intriguingly
it is at $N=6$ that the magic number $J$
sequence of $\Delta J=N$ is
broken \cite{macdonald}.
The present formalism which emphasises classical
configurations may
shed light on a possible link here.

Finally we note that the $\beta/r^2$ interaction ($\beta>0$)
is not unrealistic in quantum dots
due to the presence of image charges; in particular it
resembles the dipole-like
form
used successfully in Ref. 4. Furthermore, recent
quantitative comparisons
\cite{kinaret}\cite{madhav}\cite{review}
have
shown that the $1/r^2$ and $1/r$ repulsive interactions
yield energy spectra with very similar features (e.g.
ground state $J$ transitions, the relative excitation
$2\hbar\omega_0(B)$ for $N=2$ \cite{review});
the above results for $N=3$ are consistent with this finding.
Significant differences will only arise for the case of
attractive forces $\beta<0$ (e.g. between
electrons and holes) because of the
increased
importance of the $r\rightarrow 0$ dynamics for that case.

This work was supported by COLCIENCIAS Project No. 1204-05-264-94.

\newpage
\centerline{\bf Figure Captions}

\bigskip

\noindent Figure 1:
The $N=3$ system.
Reading clockwise, classical configuration for three
repulsive particles
(132) corresponds to
$(\alpha,\theta)=(\frac{\pi}{4},\frac{\pi}{2})$
(i.e. $(x,y)=(0,0)$);
(123) corresponds to
$(\alpha,\theta)=(\frac{\pi}{4},-\frac{\pi}{2})$
(i.e. $(x,y)=(0,\pi)$ or, equivalently,
$(0,-\pi)$).

\bigskip

\noindent Figure 2:
Contour plot of ficticious potential $ V(x,y;\epsilon)$
in the $(x,y)$ plane
for $N=3$. Relevant corresponding configurations are shown.
Minima in $V(x,y;\epsilon)$ occur at $(0,0)$ and $(0,\pm\pi)$
(i.e. at classical
configurations).
Maxima occur at $({\rm ln} {\sqrt 3}, \pm\frac{\pi}{2})$,
where
$V(x,y;\epsilon)\rightarrow\infty$ (i.e. particles 2 and
3 or 1 and 3 coincident).
$V(x,y;\epsilon)$ is positive and finite everywhere else.
The same qualitative features appear for all $\epsilon$
($\frac{\epsilon}{m^*\beta}=5$ for
illustration).


\begin{thebibliography}{99}
\bibitem{beta}
$\beta$ is chosen so that the Coulomb and $\beta/r^2$
interactions coincide for
$r\sim 150$A (i.e. typical electron separation).
Results are similar for other $\beta$ values.
\end{thebibliography}
\end{document}